\documentclass[final]{IEEEtran}

\usepackage{amssymb,amsmath}
\usepackage{array}
\usepackage{graphicx}
\usepackage{array}
\usepackage{cite}
\usepackage{stfloats}
\usepackage{subfig}
\usepackage{multirow}

\makeatletter
\def\sharrayrulewidth{\arrayrulewidth}
\def\shhline{\noalign{\ifnum0=`}\fi%
  \@ifnextchar[{\sh@hline}{\sh@hline[\sharrayrulewidth]}}
\def\sh@hline[#1]{\hrule height #1 \futurelet\reserved@a\@xhline}
\def\shvline{%
  \@ifnextchar[{\sh@vline}{\sh@vline[\sharrayrulewidth]}}
\def\sh@vline[#1]{\vrule \@width #1}
\def\shclinespace{\omit%
  \@ifnextchar[{\sh@clinespace}{\sh@clinespace[\sharrayrulewidth]}}
\def\sh@clinespace[#1]{%
  \cr
  \hbox{\vrule height \arraystretch #1 width 0pt}
}
\def\shcline{\omit%
  \@ifnextchar[{\sh@cline}{\sh@cline[\sharrayrulewidth]}}
\def\sh@cline[#1]#2{\sh@@cline[#1]#2\@nil}
\def\sh@@cline[#1]#2-#3\@nil{%
  \@multicnt#2%
  \advance\@multispan\m@ne
  \ifnum\@multicnt=\@ne\@firstofone{&\omit}\fi
  \@multicnt#3%
  \advance\@multicnt-#2%
  \advance\@multispan\@ne
  \leaders\hrule\@height#1\hfill
  \cr
  \noalign{\vskip -#1}
}

\def\fharray#1#2#3#4#5#6#7{
  \setbox\z@\hbox{$\begin{array}{#6}#7\end{array}$}
  \setbox\tw@\hbox{$\left(
    \kern-#1
    \vcenter{\kern-#3\box\z@\kern-#4}
    \kern-#2
    \right)$}
  \setbox\z@\hbox{$
    \kern#5
    \box\tw@$}
  \box\z@
}

\makeatother

\newtheorem{example}{Example}

\newcommand{\bs}[1]{\ensuremath{\boldsymbol{#1}}}
\newcommand{\sm}[1]{\text{\scriptsize \textbf{#1}}}
\providecommand{\abs}[1]{\lvert#1\rvert}
\newcommand{\amountt}{\abs{\text{\footnotesize\raisebox{0.1em}{$\sum$}}}}
\newcommand{\amountm}{\abs{\text{\tiny\raisebox{0.25em}{$\sum$}}}}
\newcommand{\figref}[1]{Fig.~\ref{fig:#1}}
\newcommand{\secref}[1]{Section~\ref{sec:#1}}
\newcommand{\tabref}[1]{Table~\ref{tab:#1}}
\newcommand{\dependent}{\;\middle\vert\;}

\begin{document}

\title{A Closed Form Expression for the\\Exact Bit Error Probability for Viterbi Decoding\\of Convolutional Codes}

\author{
  \IEEEauthorblockN{Irina E. Bocharova, Florian Hug,~\IEEEmembership{Student Member,~IEEE},\\Rolf Johannesson,~\IEEEmembership{Life~Fellow,~IEEE}, and Boris D. Kudryashov}
  \thanks{This work was supported in part by the Swedish Research Council by Grant 621-2007-6281.}%
  \thanks{The material in this paper was presented in part at the Information Theory and Applications Workshop, San Diego, USA, 2011 and the International Mathematical Conference ``50 Years Of IPPI'', Moscow, Russia, 2011.}%
  \thanks{I. E. Bocharova and B. D. Kudryashov are with the Department of Information Systems, St. Petersburg University of Information Technologies, Mechanics and Optics, St. Petersburg 197101, Russia (e-mail: irina@eit.lth.se; boris@eit.lth.se).}%
  \thanks{F. Hug and R. Johannesson are with the Department of Electrical and Information Technology, Lund University, SE-22100 Lund, Sweden (e-mail: florian@eit.lth.se; rolf@eit.lth.se).}%
  \vspace{-1em}%
}

\maketitle
\begin{abstract}
  In 1995, Best et al. published a formula for the exact bit error probability for Viterbi decoding of the rate $R=1/2$, memory $m=1$ ($2$-state) convolutional encoder with generator matrix $G(D)=(1 \quad 1+D)$ when used to communicate over the binary symmetric channel. Their formula was later extended to the rate $R=1/2$, memory $m=2$ ($4$-state) convolutional encoder with generator matrix $G(D)=(1+D^2 \quad 1+D+D^2)$ by Lentmaier et al.

  In this paper, a different approach to derive the exact bit error probability is described. A general recurrent matrix equation, connecting the average information weight at the current and previous states of a trellis section of the Viterbi decoder, is derived and solved. The general solution of this matrix equation yields a closed form expression for the exact bit error probability. As special cases, the expressions obtained by Best et al. for the $2$-state encoder and by Lentmaier et al. for a $4$-state encoder are obtained. The closed form expression derived in this paper is evaluated for various realizations of encoders, including rate $R=1/2$ and $R=2/3$ encoders, of as many as $16$ states.

  Moreover, it is shown that it is straightforward to extend the approach to communication over the quantized additive white Gaussian noise channel.
\end{abstract}


\section{Introduction}
In 1971, Viterbi \cite{Viterbi1971} published a nowadays classical upper bound on the bit error probability $P_\text{b}$ for Viterbi decoding, when convolutional codes are used to communicate over the binary symmetric channel (BSC). This bound was derived from the extended path weight enumerators, obtained using a signal flow chart technique for convolutional encoders. Later, van de Meeberg \cite{Meeberg1974} used a very clever observation to tighten Viterbi's bound for large signal-to-noise ratios (SNRs).

The challenging problem of deriving an expression for the \textit{exact} (decoding) bit error probability was first addressed by Morrissey in 1970 \cite{Morrissey1970} for a suboptimal feedback decoding algorithm. He obtained the same expression for the exact bit error probability for the rate $R=1/2$, memory $m=1$ ($2$-state) convolutional encoder with generator matrix $G(D) = (1 \quad 1+D)$ that Best et al. \cite{Best1995} obtained for Viterbi decoding. Their method is based on considering a Markov chain of the so-called metric states of the Viterbi decoder; an approach due to Burnashev and Cohn \cite{Burnashev1990}. An extension of this method to the rate $R=1/2$ memory $m=2$ ($4$-state) convolutional encoder with generator matrix $G(D) = (1+D^2 \quad 1+D+D^2)$ was published by Lentmaier et al. \cite{Lentmaier2004}.

In this paper we use a different and more general approach to derive a closed form expression for the exact (decoding) bit error probability for Viterbi decoding of convolutional encoders, when communicating over the BSC as well as the quantized additive white Gaussian noise (AWGN) channel. Our new method allows the calculation of the exact bit error probability for more complex encoders in a wider range of code rates than the methods of \cite{Best1995} and \cite{Lentmaier2004}. By considering a random tie-breaking strategy, we average the information weights over the channel noise sequence and the sequence of random decisions based on coin-flippings (where the coin may have more than two sides depending on the code rate). Unlike the backward recursion in \cite{Best1995} and \cite{Lentmaier2004}, the bit error probability averaged over time is obtained by deriving and solving a recurrent matrix equation for the average information weights at the current and previous states of a trellis section when the maximum-likelihood branches are decided by the Viterbi decoder at the current step.

To illustrate our method, we use a rate $R=2/3$ systematic convolutional $2$-state encoder whose minimal realization is given in observer canonical form, since this encoder is both general and simple.

In \secref{recurrent_equation}, the problem of computing the exact bit error probability is reformulated via the average information weights. A recurrent matrix equation for these average information weights is derived in \secref{inf_weights} and solved in \secref{solving_the_recurrent_equation}. In \secref{examples}, we give additional examples of rate $R=1/2$ and $R=2/3$ encoders of various memories. Furthermore, we analyze a rate $R=1/2$ $4$-state encoder used to communicate over the quantized additive white Gaussian noise (AWGN) channel and show an interesting result that would be difficult to obtain without being able to calculate the exact bit error probability.

Before proceeding, we would like to emphasize that the bit error probability is an encoder property, neither a generator matrix property nor a convolutional code property.

\section{Problem Formulation via the Average Information Weights}\label{sec:recurrent_equation}
Assume that the all-zero sequence is transmitted over a BSC with crossover probability $p$ and let $W_t(\sigma)$ denote the weight of the information sequence corresponding to the code sequence decided by the Viterbi decoder at state $\sigma$ and time instant $t$. If the initial values $W_0(\sigma)$ are known, then the random process $W_t(\sigma)$, $t=0,1,2,\ldots$, is a function of the random sequence of the received $c$-tuples $\bs{r}_\tau$, $\tau=0,1,\ldots,t-1$, and the coin-flippings used to resolve ties.

Our goal is to determine the mathematical expectation of the random variable $W_t(\sigma)$ over this ensemble, since for rate $R=b/c$ minimal convolutional encoders the bit error probability can be computed as the limit
\begin{IEEEeqnarray}{rCl}
	\label{eq: viterbi_bit_error_probability}
	P_{\text{b}} & = & \lim_{t \to \infty} \frac{E\left[ W_t(\sigma = 0)\right]}{tb}
\end{IEEEeqnarray}
assuming that this limit exists.

\textit{Remark.} If we consider nonminimal encoders, all states equivalent to the all-zero state have to be also taken into account.

We consider encoder realizations in both controller and observer canonical form and denote the encoder states by $\sigma$, $\sigma \in \{ 0, 1, \ldots, \amountt-1 \}$, where {\footnotesize\raisebox{0.1em}{$\sum$}} is the set of all possible encoder states.

During the decoding step at time instant $t+1$ the Viterbi algorithm computes the cumulative Viterbi branch metric vector $\bs{\mu}_{t+1} = \left( \mu_{t+1}(0) \enspace \mu_{t+1}(1) \ldots \mu_{t+1}(\amountt-1) \right)$ for the time instant $t+1$ using the vector $\bs{\mu}_t$ and the received $c$-tuple $\bs{r}_t$. It is convenient to normalize the metrics such that the cumulative metrics at every all-zero state will be zero, that is, we subtract the value $\mu_t(0)$ from $\mu_t(1), \mu_t(2), \ldots, \mu_t(\amountt-1)$ and introduce the normalized cumulative branch metric vector
\begin{IEEEeqnarray}{rCl}
	\bs{\phi}_t & = & \Big( \phi_t(1) \enspace \phi_t(2) \ldots \phi_t(\amountm-1) \Big) \nonumber \\
	            & = & \Big( \mu_t(1)\!-\!\mu_t(0) \enspace \mu_t(2)\!-\!\mu_t(0) \ldots \mu_t(\amountm-1)\!-\!\mu_t(0) \Big) \nonumber
\end{IEEEeqnarray}
For example, for a $2$-state encoder we obtain the scalar
\begin{IEEEeqnarray}{rCl}
	\phi_t & = & \phi_t(1) \nonumber
\end{IEEEeqnarray}
while for a $4$-state encoder we have the vector
\begin{IEEEeqnarray}{rCl}
	\bs{\phi}_t & = & \Big( \phi_t(1) \enspace \phi_t(2) \enspace \phi_t(3) \Big) \nonumber
\end{IEEEeqnarray}

The elements of the random vector $\bs{\phi}_t$ belong to a set whose cardinality $M$ depends on the channel model, encoder structure, and the tie-breaking rule. Enumerating the vectors $\bs{\phi}_t$ by numbers $\phi_{t}$ which are random variables taking on $M$ different integer values $\phi^{(0)},\,\phi^{(1)},\ldots,\,\phi^{(M-1)}$, the sequence of numbers $\phi_{t}$ forms an $M$-state Markov chain $\Phi_{t}$  with transition probability matrix $\Phi = \left( \phi_{jk} \right)$, where
\begin{equation}
  \phi_{jk} = \text{Pr} \left( \phi_{t+1} = \phi^{(k)} \dependent \phi_t = \phi^{(j)} \right)
\end{equation}

\begin{figure}[t]
  \centering
  \includegraphics{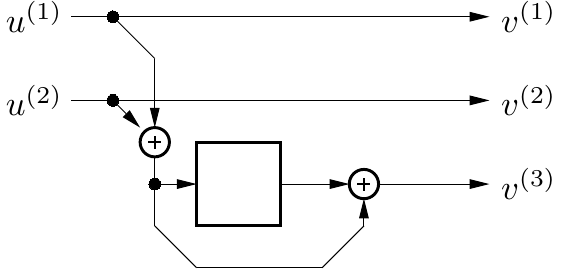}
  \caption{\label{fig:ocf_realization}A minimal encoder for the generator matrix given in equation \eqref{eq:ocf_encoding_matrix_R23}.}
\end{figure}

\begin{figure*}[t]
  \subfloat[]{\hspace{-15.9339pt}\includegraphics{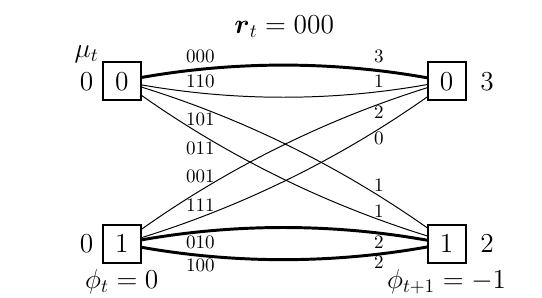}\hspace{-14.2272pt}}
  \subfloat[]{\hspace{-15.9339pt}\includegraphics{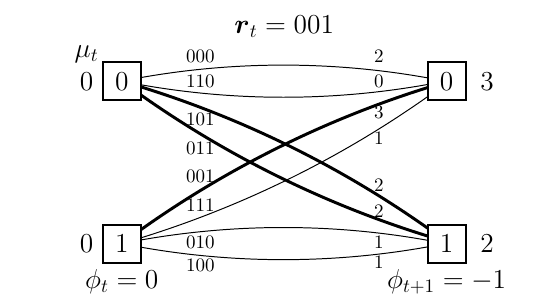}\hspace{-14.2272pt}}
  \subfloat[]{\hspace{-15.9339pt}\includegraphics{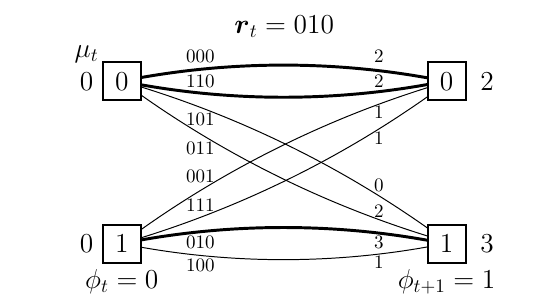}\hspace{-14.2272pt}}
  \subfloat[]{\hspace{-15.9339pt}\includegraphics{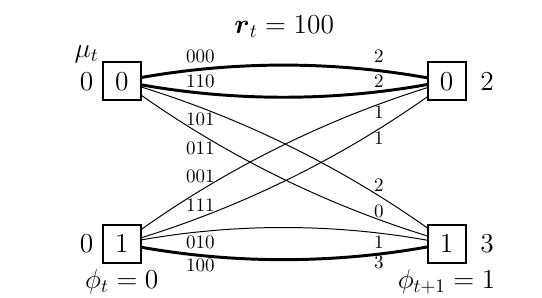}\hspace{-14.2272pt}} \\[-2mm]
  \subfloat[]{\hspace{-15.9339pt}\includegraphics{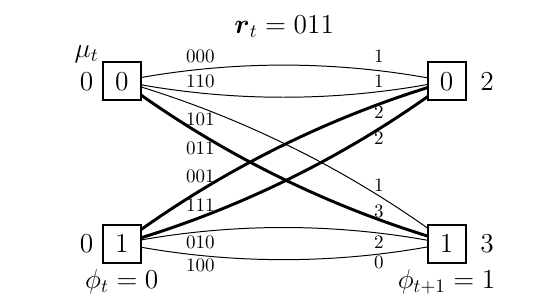}\hspace{-14.2272pt}}
  \subfloat[]{\hspace{-15.9339pt}\includegraphics{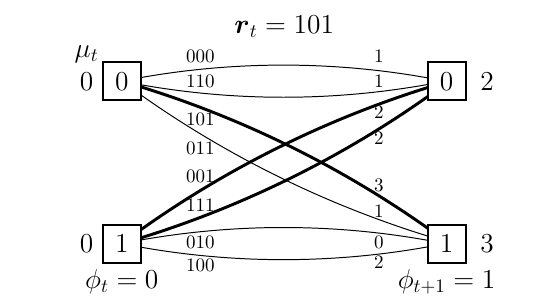}\hspace{-14.2272pt}}
  \subfloat[]{\hspace{-15.9339pt}\includegraphics{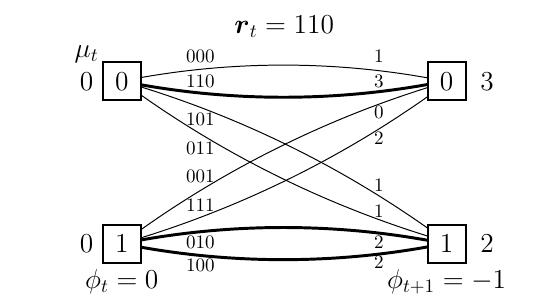}\hspace{-14.2272pt}}
  \subfloat[]{\hspace{-15.9339pt}\includegraphics{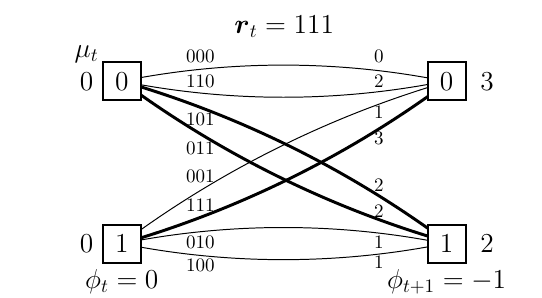}\hspace{-14.2272pt}}
  \caption{\label{fig:eight_trellis_sections}Eight (of a total of $40$) trellis sections for the rate $R=2/3$, $2$-state encoder in \figref{ocf_realization}.}
\end{figure*}

Let $\bs{W}_t$ be the vector of information weights at time instant $t$ that depends both on the $\amountt$ encoder states $\sigma_t$ and on the $M$ normalized cumulative metrics $\phi_t$; that is, $\bs{W}_t$ is expressed as the following vector with $M\amountt$ entries
\begin{IEEEeqnarray}{rclccccr}
	\bs{W}_t & = & \Big( & \bs{W}_t(\sigma=0) \enspace & \bs{W}_t(\sigma=1) & \enspace \ldots \enspace & \bs{W}_t(\sigma=\amountm\!-\!1) \Big) \IEEEeqnarraynumspace
\end{IEEEeqnarray}
where
\begin{IEEEeqnarray}{rclcccr}
	\bs{W}_t(\sigma) & = & \Big( & W_t(\phi^{(0)}, \sigma) \enspace & W_t(\phi^{(1)}, \sigma)  & \enspace \ldots \enspace & W_t(\phi^{(M-1)}, \sigma) \Big) \IEEEeqnarraynumspace
\end{IEEEeqnarray}
Then \eqref{eq: viterbi_bit_error_probability} can be rewritten as
\begin{IEEEeqnarray}{rCl}
	P_{\text{b}} & = & \lim_{t \to \infty} \frac{E[W_t(\sigma = 0)]}{tb} = \lim_{t \to \infty} \frac{\sum_{i=0}^{M\!-\!1} E[W_t(\phi^{(i)}, \sigma = 0)]}{tb} \nonumber \\[0.5mm]
	    & = & \lim_{t \to \infty} \frac{E[\bs{W}_t(\sigma = 0)] \bs{1}^\text{T}_{1,M}}{tb} = \lim_{t \to \infty} \frac{\bs{w}_t(\sigma = 0) \bs{1}^\text{T}_{1,M}}{tb}\nonumber \\[0.5mm]
        & = & \lim_{t \to \infty} \frac{\bs{w}_t}{tb} \big( \bs{1}_{1,M}\, \bs{0}_{1,M}\ldots\,\bs{0}_{1,M})^{\text{T}} \label{eq: viterbi_bit_error_probability_w}
 \end{IEEEeqnarray}
where $\bs{1}_{1,M}$ and $\bs{0}_{1,M}$ denote the all-one and the all-zero row vectors of length $M$, respectively, $\bs{w}_{t}$ represents the length $M \amountt$ vector of the average information weights, while the length $M$ vector of average information weights at the state $\sigma$ is given by $\bs{w}_t(\sigma)$. Note that the mathematical expectations in \eqref{eq: viterbi_bit_error_probability_w} are computed over the sequences of channel noises and coin-flipping decisions.

To illustrate the introduced notations, we use the rate $R=2/3$ memory $m=1$, overall constraint length $\nu=2$, minimal encoder with systematic generator matrix
\begin{equation}
  G(D) =
  \left(\begin{array}{ccc}
    1 & 0 & 1+D \\
    0 & 1 & 1+D
  \end{array}\right) \label{eq:ocf_encoding_matrix_R23}
\end{equation}
It has a $2$-state realization in observer canonical form as shown in \figref{ocf_realization}.

\begin{figure}[t]
	\centering
	\includegraphics{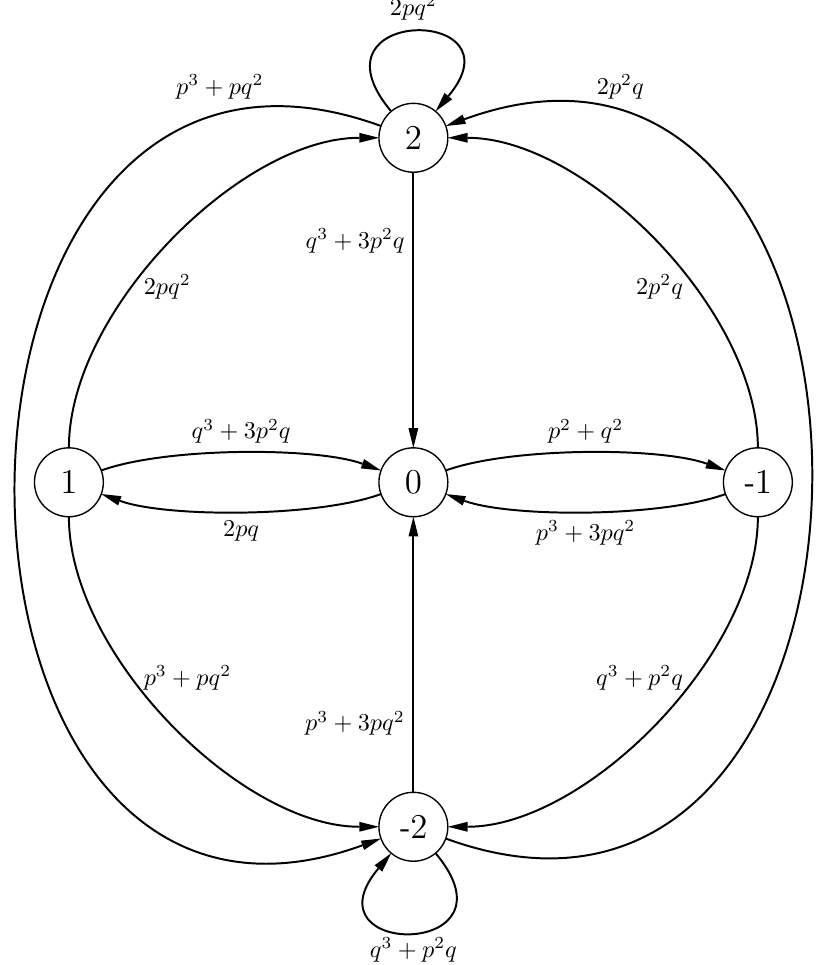}
	\caption{\label{fig:state_transition_diagram}Illustration of the $5$-state Markov chain formed by the sequences of normalized cumulative metric states $\phi_t$.}
	\vspace{-1em}
\end{figure}

Assuming that the normalized cumulative metric state is $\phi_t = 0$, we obtain the eight trellis sections given in \figref{eight_trellis_sections}. These trellis sections yield the normalized cumulative metric states $\left\{ -1, 0, 1 \right\}$. Using $\phi_t = -1$ and $\phi_t = 1$, we obtain $16$ additional trellis sections and two additional normalized cumulative metric states $\left\{ -2, 2 \right\}$. From the metrics $\phi_t = -2$ and $\phi_t = 2$, we get another $16$ trellis sections but those will not yield any new metrics. Thus, in total we have $M=5$ normalized cumulative metric states $\phi_t \in \left\{ -2, -1, 0, 1, 2\right\}$. Together with the eight different received triples, $\bs{r}_t = 000$, $001$, $010$, $100$, $011$, $101$, $110$, $111$, they correspond to in total $40$ different trellis sections. The bold branches in \figref{eight_trellis_sections} correspond to the branches decided by the Viterbi decoder at time instant $t+1$. When we have more than one branch with maximum normalized cumulative metric entering the same state, we have a tie which we, in our analysis, resolve by fair coin-flipping.

Hence, the normalized cumulative metric $\Phi_t$ is a $5$-state Markov chain with transition probability matrix $\Phi = \left( \phi_{jk} \right)$, $1 \leq j,k \leq 5$.

From the four trellis sections, (a), (b), (g), and (h), in \figref{eight_trellis_sections} we obtain
\begin{IEEEeqnarray}{rCl}
  \phi_{0(-1)} & = & \text{Pr}\left( \bs{r}_t = 000 \right) + \text{Pr}\left( \bs{r}_t = 001 \right) \nonumber \\
  &   & \,+\,\text{Pr}\left( \bs{r}_t = 110 \right) + \text{Pr}\left( \bs{r}_t = 111 \right) \nonumber \\
  & = &  q^3 + pq^2 + p^2 q + p^3 = p^2 + q^2
\end{IEEEeqnarray}
while the four trellis sections, (c), (d), (e), and (f), yield
\begin{equation}
  \phi_{01} = pq^2 + pq^2 + p^2q + p^2q = 2pq
\end{equation}
where $q=1-p$.

Similarly, we can obtain the remaining transition probabilities from the $32$ trellis sections not included in \figref{eight_trellis_sections}. Their transition probability matrix follows as
\vspace{1em}
\begin{IEEEeqnarray}{c}
   \Phi = \fharray{2em}{2em}{1em}{1em}{1em}{l@{\hspace{0em}}ccccc@{\hspace{0em}}c}{
                     & \sm{-2}    & \sm{-1}   & \sm{0}      & \sm{1} & \sm{2} & \phi^{(k)}\\
   \sm{-2}           & q^3 + p^2q & 0         & p^3 + 3pq^2 & 0      & 2p^2q \\
   \sm{-1}           & q^3 + p^2q & 0         & p^3 + 3pq^2 & 0      & 2p^2q \\
   \sm{\phantom{-}0} & 0          & p^2 + q^2 & 0           & 2pq    & 0     \\
   \sm{\phantom{-}1} & p^3 + pq^2 & 0         & q^3 + 3p^2q & 0      & 2p^2q \\
   \sm{\phantom{-}2} & p^3 + pq^2 & 0         & q^3 + 3p^2q & 0      & 2p^2q \\
   \phi^{(j)}
   }\vspace{1em}\IEEEeqnarraynumspace
\end{IEEEeqnarray}
whose metric state Markov chain is shown in \figref{state_transition_diagram}.

Let $\bs{p}_t$ denote the probabilities of the $M$ different normalized cumulative metric values of $\Phi_t$, that is, $\phi_t \in \{ \phi^{(0)}, \phi^{(1)}, \ldots, \phi^{(M-1)} \}$. Their stationary distribution is denoted $\bs{p}_{\infty} = ( p_{\infty}^{(0)} \enspace p_{\infty}^{(1)} \ldots p_{\infty}^{(M-1)} )$ and is determined as the solution of, for example, the first $M-1$ equations of
\begin{IEEEeqnarray}{rCl}
\label{steady-state}
	\bs{p}_{\infty} \Phi & = & \bs{p}_{\infty}
\end{IEEEeqnarray}
and
\vspace{-1em}
\begin{IEEEeqnarray}{rCl}
	\sum_{i=0}^{M-1} p_{\infty}^{(i)} & = & 1
\end{IEEEeqnarray}
For the $2$-state convolutional encoder with generator matrix \eqref{eq:ocf_encoding_matrix_R23} we obtain
{\setlength\arraycolsep{0.1em}
\renewcommand{\arraystretch}{1.2}
\small
\begin{IEEEeqnarray*}{c}
	\bs{p}_{\infty}^\text{T} = \frac{1}{1 -p +10p^2 -20p^3 +20p^4 -8p^5} \\
	\times \left(\begin{array}{ccccccccccccccc}
	  1 & + & 7p & - & 28p^2 & + & 66p^3 & - & 100p^4 & + & 96p^5 & - & 56p^6 & + & 16p^7 \\
	    & - & 3p & + & 16p^2 & - & 46p^3 & + &  80p^4 & - & 88p^5 & + & 56p^6 & - & 16p^7 \\
	    & - & 3p & + & 10p^2 & - & 20p^3 & + &  20p^4 & - &  8p^5 \\
	    &   &    & - &  6p^2 & + & 26p^3 & - &  60p^4 & + & 80p^5 & - & 56p^6 & - & 16p^7 \\
	    &   &    & - &  2p^2 & - &  6p^3 & + &  40p^4 & - & 72p^5 & + & 56p^6 & - & 16p^7
	\end{array}\right)
\end{IEEEeqnarray*}}%
In order to compute the exact bit error probability according to \eqref{eq: viterbi_bit_error_probability_w}, it is necessary to determine $\bs{w}_{t}(\sigma=0)$. In the next section we will derive a recurrent matrix equation for the average information weights and illustrate how to obtain its components using as an example the rate $R=2/3$ memory $m=1$ minimal encoder determined by \eqref{eq:ocf_encoding_matrix_R23}.

\section{Computing the Vector of Average Information Weights}\label{sec:inf_weights}
The vector $\bs{w}_t$ describes the dynamics of the information weights when we proceed along the trellis and satisfies the recurrent equation
\begin{IEEEeqnarray}{C}
	\label{eq: recurrent_equation}
	\left\{ \,
	\begin{IEEEeqnarraybox}[\IEEEeqnarraystrutmode\IEEEeqnarraystrutsizeadd{2pt}{2pt}][c]{rCl}
		\bs{w}_{t+1} & = & \bs{w}_t A + \bs{b}_t B \\
		\bs{b}_{t+1} & = & \bs{b}_t \Pi
	\end{IEEEeqnarraybox} \right.
\end{IEEEeqnarray}
where $A$ and $B$ are $M\amountt \times M\amountt$ nonnegative matrices, $\Pi$ is an $M\amountt \times M\amountt$ stochastic matrix, and $\amountt=2^m$. Both matrices consist of $\amountt \times \amountt$ submatrices $A_{ij}$ and $B_{ij}$ of size $M \times M$, respectively, where the former satisfy
 \begin{equation}
  \sum_{i=0}^{\amountm-1} A_{ij} = {\Phi-1}, \enspace j=0,1,\ldots,\amountm-1 \label{eq:row_sum}
\end{equation}
since we consider only encoders for which every encoder state is reachable with probability $1$.

The matrix A represents the linear part of the affine transformation of the information weights while the matrix $B$ describes their increments. The submatrices $A_{ij}$ and $B_{ij}$ describe the updating of the average information weights if the transition from state $i$ to state $j$ exists; and are zero otherwise. Moreover, the vector $\bs{b}_t$ of length $M\amountt$ is the concatenation of $\amountt$ stochastic vectors $\bs{p}_t$, and hence the $M\amountt \times M\amountt$ matrix~$\Pi$ follows as
\begin{IEEEeqnarray}{rCl}
	\Pi = \left(\begin{array}{cccc}
		\Phi   & 0      & \ldots & 0 \\
		0      & \Phi   & \ldots & 0 \\
		\vdots & \vdots & \ddots & \vdots \\
		0      & 0      & \ldots & \Phi
	\end{array}\right)
\end{IEEEeqnarray}
For simplicity, we choose the initial value of the vector of the information weights to be
\begin{IEEEeqnarray}{rCl}
	\label{eq: initial_value_w}
	\bs{w}_0 & = & \bs{0}
\end{IEEEeqnarray}

Continuing the previous example, we will illustrate how the $10 \times 10$ matrices $A$ and $B$ can be obtained directly from all $40$ trellis sections. For example, the eight trellis sections in \figref{eight_trellis_sections} determine all transitions from $\phi_t = 0$ to either $\phi_{t+1} = -1$ or $\phi_{t+1} = 1$.

To be more specific, consider all transitions from $\sigma_t = 0$ and $\phi_t = 0$ to $\sigma_{t+1} = 0$ and $\phi_{t+1} = -1$, as shown in \figref{eight_trellis_sections}(a), (b), (g), and (h). Only \figref{eight_trellis_sections}(a) and (g) have transitions decided by the Viterbi algorithm, which are $\bs{v}_t = 000$ in \figref{eight_trellis_sections}(a) and $\bs{v}_t = 110$ in \figref{eight_trellis_sections}(g), and thus the entry $\sigma_t = 0$, $\phi_t = 0$, $\sigma_{t+1} = 0$, $\phi_{t+1} = -1$ in matrix $A$ follows as
\begin{equation}
  \text{Pr}\left( \bs{r}_t = 000 \right) + \text{Pr}\left( \bs{r}_t = 110 \right) = q^3 + p^2q \nonumber
\end{equation}
and in matrix $B$ as
\begin{IEEEeqnarray}{rCl}
  \IEEEeqnarraymulticol{3}{l}{\beta\left(000\right)\text{Pr}\left( \bs{r}_t = 000 \right) + \beta\left(110\right)\text{Pr}\left( \bs{r}_t = 110 \right)} \nonumber \\
  \quad & = & 0 + 2p^2q = 2p^2q \nonumber
\end{IEEEeqnarray}
where $\beta\left( \bs{v}_t \right)$ denotes the number of information $1$s corresponding to $\bs{v}_t$. Since we use coin-flipping to resolve ties, we obtain that the entry $\sigma_t = 0$, $\phi_t = 0$, $\sigma_{t+1} = 0$, $\phi_{t+1} = 1$ (\figref{eight_trellis_sections}(c) and (d)) in matrix $A$ is
\begin{IEEEeqnarray}{rCl}
  \IEEEeqnarraymulticol{3}{l}{\frac{1}{2} \text{Pr} \left( \bs{r}_t = 010 \right) + \frac{1}{2} \text{Pr} \left( \bs{r}_t = 010 \right)} \nonumber \\
  \IEEEeqnarraymulticol{3}{l}{\, + \frac{1}{2} \text{Pr} \left( \bs{r}_t = 100 \right) + \frac{1}{2} \text{Pr} \left( \bs{r}_t = 100 \right)} \nonumber \\
  \quad & = & \frac{1}{2} pq^2 + \frac{1}{2} pq^2 + \frac{1}{2} pq^2 + \frac{1}{2} pq^2 = 2pq^2 \nonumber
\end{IEEEeqnarray}
and in matrix $B$
\begin{IEEEeqnarray}{rCl}
  \IEEEeqnarraymulticol{3}{l}{\frac{1}{2} \beta\left(000\right)\text{Pr} \left( \bs{r}_t = 010 \right) + \frac{1}{2} \beta\left(110\right) \text{Pr} \left( \bs{r}_t = 010 \right)} \nonumber \\
  \IEEEeqnarraymulticol{3}{l}{\, + \frac{1}{2} \beta\left(000\right) \text{Pr} \left( \bs{r}_t = 100 \right) + \frac{1}{2} \beta\left(110\right) \text{Pr} \left( \bs{r}_t = 100 \right) } \nonumber \\
  \quad & = & \frac{1}{2} \cdot 0 + \frac{1}{2} \cdot 2 pq^2 + \frac{1}{2} \cdot 0 + \frac{1}{2} \cdot 2 pq^2 = 2pq^2 \nonumber
\end{IEEEeqnarray}
Similarly the entry $\sigma_t = 1$, $\phi_t = 0$, $\sigma_{t+1} = 0$, $\phi_{t+1} = -1$ (\figref{eight_trellis_sections}(b) and (h)) in matrix $A$ is
\begin{equation*}
  pq^2 + p^3
\end{equation*}
and in matrix $B$
\begin{equation*}
  0 + 2p^3 = 2p^3
\end{equation*}
Finally, the entry $\sigma_t = 1$, $\phi_t =0$, $\sigma_{t+1} = 0$, $\phi_{t+1} = 1$ (\figref{eight_trellis_sections}(e) and (f)) in matrix $A$ is given by
\begin{equation*}
  \frac{1}{2} p^2 q + \frac{1}{2} p^2q + \frac{1}{2} p^2 q + \frac{1}{2} p^2 q = 2p^2q
\end{equation*}
and in matrix $B$ by
\begin{equation*}
  \frac{1}{2} \cdot 0 + \frac{1}{2} \cdot 2 p^2 q + \frac{1}{2} \cdot 0 + \frac{1}{2} \cdot 2 p^2 q = 2p^2q
\end{equation*}

The trellis sections in \figref{eight_trellis_sections} determine also the entries for the transitions $\sigma_t = 0$, $\phi_t = 0$, $\sigma_{t+1} = 1$, $\phi_{t+1} = -1$ and $\sigma_t = 0$, $\phi_t = 0$, $\sigma_{t+1} = 1$, $\phi_{t+1} = 1$ as well as the transitions $\sigma_t =1$, $\phi_t = 0$, $\sigma_{t+1} = 1$, $\phi_{t+1} = -1$ and $\sigma_t = 1$, $\phi_t = 0$, $\sigma_{t+1} = 1$, $\phi_{t+1} = 1$.

The remaining transitions with $\phi_t = 0$ are never decided by the Viterbi algorithm, and hence the corresponding entries are zero. The eight trellis sections in \figref{eight_trellis_sections} yield $20$ entries in the matrices $A$ and $B$, while the $32$ trellis sections not shown in \figref{eight_trellis_sections} yield the remaining $80$ entries. For the convolutional encoder shown in \figref{ocf_realization} we obtain
\begin{equation}
  A = \left( \begin{array}{cc} A_{00} & A_{01} \\ A_{10} & A_{11} \end{array} \right)
\end{equation}
where
\vspace{1em}
\begin{IEEEeqnarray}{c}
	\setlength\arraycolsep{0.2em}
	\renewcommand{\arraystretch}{1.1}
	A_{00} = \fharray{2em}{0em}{1em}{0em}{1em}{r@{\hspace{1.5em}}ccccc}{
		        & \sm{-2}      &\sm{-1}       & \sm{0}                             & \sm{1} & \sm{2} \\
		\sm{-2} & q^3\!+\!p^2q & 0            & p^3\!+\!3pq^2                      & 0      & 2p^2q \\
		\sm{-1} & q^3\!+\!p^2q & 0            & \frac{1}{2}p^3\!+\!\frac{5}{2}pq^2 & 0      & p^2q  \\
		\sm{0}  & 0            & q^3\!+\!p^2q & 0                                  & 2pq^2  & 0 \\
		\sm{1}  & 0            & 0            & \frac{1}{2}q^3\!+\!\frac{1}{2}p^2q & 0      & pq^2  \\
		\sm{2}  & 0            & 0            & 0                                  & 0      & 0
	} \IEEEeqnarraynumspace
\end{IEEEeqnarray}
\vspace{1em}
\begin{IEEEeqnarray}{c}
	\setlength\arraycolsep{0.2em}
	\renewcommand{\arraystretch}{1.1}
	A_{01} = \fharray{2em}{0em}{1em}{0em}{1em}{r@{\hspace{1.5em}}ccccc}{
		        & \sm{-2}      &\sm{-1}       & \sm{0}                             & \sm{1} & \sm{2} \\
		\sm{-2} & p^2q\!+\!q^3                       & 0            & p^3\!+\!3pq^2 & 0     & 2p^2q \\
		\sm{-1} & \frac{1}{2}p^3\!+\!\frac{1}{2}pq^2 & 0            & p^2q          & 0     & 0  \\
		\sm{0}  & 0                                  & p^3\!+\!pq^2 & 0             & 2p^2q & 0 \\
		\sm{1}  & \frac{1}{2}q^3\!+\!\frac{1}{2}p^2q & 0            & p^3\!+\!2pq^2 & 0     & 2p^2q  \\
		\sm{2}  & 0                                  & 0            & 0             & 0     & 0
	} \IEEEeqnarraynumspace
\end{IEEEeqnarray}
\vspace{1em}
\begin{IEEEeqnarray}{c}
	\setlength\arraycolsep{0.2em}
	\renewcommand{\arraystretch}{1.1}
	A_{10} = \fharray{2em}{0em}{1em}{0em}{1em}{r@{\hspace{1.5em}}ccccc}{
		        & \sm{-2}      &\sm{-1}       & \sm{0}                             & \sm{1} & \sm{2} \\
		\sm{-2} & 0            & 0            & 0                                  & 0      & 0 \\
		\sm{-1} & 0            & 0            & \frac{1}{2}p^3\!+\!\frac{1}{2}pq^2 & 0      & p^2q  \\
		\sm{0}  & 0            & p^3\!+\!pq^2 & 0                                  & 2p^2q  & 0 \\
		\sm{1}  & p^3\!+\!pq^2 & 0            & \frac{1}{2}q^3\!+\!\frac{5}{2}p^2q & 0      & pq^2  \\
		\sm{2}  & p^3\!+\!pq^2 & 0            & 3p^2q\!+\!q^3                      & 0      & 2pq^2
	} \IEEEeqnarraynumspace
\end{IEEEeqnarray}
\vspace{1em}
\begin{IEEEeqnarray}{c}
	\setlength\arraycolsep{0.2em}
	\renewcommand{\arraystretch}{1.1}
	A_{11} = \fharray{2em}{0em}{1em}{0em}{1em}{r@{\hspace{1.5em}}ccccc}{
		        & \sm{-2}      &\sm{-1}       & \sm{0}                             & \sm{1} & \sm{2} \\
		\sm{-2} & 0                                  & 0            & 0             & 0      & 0      \\
		\sm{-1} & \frac{1}{2}p^2q\!+\!\frac{1}{2}q^3 & 0            & pq^2          & 0      & 0      \\
		\sm{0}  & 0                                  & p^2q\!+\!q^3 & 0             & 2pq^2  & 0      \\
		\sm{1}  & \frac{1}{2}pq^2\!+\!\frac{1}{2}p^3 & 0            & 2p^2q\!+\!q^3 & 0      & 2pq^2  \\
		\sm{2}  & p^3\!+\!pq^2                       & 0            & 3p^2q\!+\!q^3 & 0      & 2pq^2
	} \IEEEeqnarraynumspace
\end{IEEEeqnarray}
and
\begin{equation}
  B = \left( \begin{array}{cc} B_{00} & B_{01} \\ B_{10} & B_{11} \end{array} \right)
\end{equation}
where
\vspace{1em}
\begin{IEEEeqnarray}{c}
	\setlength\arraycolsep{0.2em}
	\renewcommand{\arraystretch}{1.1}
	B_{00} = \fharray{2em}{0em}{1em}{0em}{1em}{r@{\hspace{1.5em}}ccccc}{
		        & \sm{-2} &\sm{-1}  & \sm{0}         & \sm{1} & \sm{2} \\
		\sm{-2} & 2p^2q   & 0       & p^3\!+\!2pq^2 & 0      & 2p^2q \\
		\sm{-1} & 0       & 0       & p^2q           & 0      & pq^2  \\
		\sm{0}  & 0       & 2p^2q   & 0              & 2pq^2  & 0 \\
		\sm{1}  & 2p^2q   & 0       & p^3\!+\!2pq^2  & 0      & p^2q  \\
		\sm{2}  & 0       & 0       & 0              & 0      & 0
	} \IEEEeqnarraynumspace
\end{IEEEeqnarray}
\vspace{1em}
\begin{IEEEeqnarray}{c}
	\setlength\arraycolsep{0.2em}
	\renewcommand{\arraystretch}{1.1}
	B_{01} = \fharray{2em}{0em}{1em}{0em}{1em}{r@{\hspace{1.5em}}ccccc}{
		        & \sm{-2}                            &\sm{-1}       & \sm{0}        & \sm{1} & \sm{2} \\
		\sm{-2} & p^2q\!+\!q^3                       & 0            & p^3\!+\!3pq^2 & 0      & 2p^2q \\
		\sm{-1} & \frac{1}{2}p^3\!+\!\frac{1}{2}pq^2 & 0            & p^2q          & 0      & 0  \\
		\sm{0}  & 0                                  & p^3\!+\!pq^2 & 0             & 2p^2q  & 0 \\
		\sm{1}  & \frac{1}{2}q^3\!+\!\frac{1}{2}p^2q & 0            & p^3\!+\!2pq^2 & 0      & 2p^2q  \\
		\sm{2}  & 0                                  & 0            & 0             & 0      & 0
	} \IEEEeqnarraynumspace
\end{IEEEeqnarray}
\vspace{1em}
\begin{IEEEeqnarray}{c}
	\setlength\arraycolsep{0.2em}
	\renewcommand{\arraystretch}{1.1}
	B_{10} = \fharray{2em}{0em}{1em}{0em}{1em}{r@{\hspace{1.5em}}ccccc}{
		        & \sm{-2} &\sm{-1} & \sm{0} & \sm{1} & \sm{2} \\
		\sm{-2} & 0       & 0      & 0      & 0      & 0 \\
		\sm{-1} & 0       & 0      & p^3    & 0      & p^2q  \\
		\sm{0}  & 0       & 2p^3   & 0      & 2p^2q  & 0 \\
		\sm{1}  & 2p^3    & 0      & 3p^2q  & 0      & pq^2  \\
		\sm{2}  & 2p^3    & 0      & 4p^2q  & 0      & 2pq^2
	} \IEEEeqnarraynumspace
\end{IEEEeqnarray}
\vspace{1em}
\begin{IEEEeqnarray}{c}
	\setlength\arraycolsep{0.2em}
	\renewcommand{\arraystretch}{1.1}
	B_{11} = \fharray{2em}{0em}{1em}{0em}{1em}{r@{\hspace{1.5em}}ccccc}{
		        & \sm{-2}                            &\sm{-1}       & \sm{0}        & \sm{1} & \sm{2} \\
		\sm{-2} & 0                                  & 0            & 0             & 0      & 0      \\
		\sm{-1} & \frac{1}{2}p^2q\!+\!\frac{1}{2}q^3 & 0            & pq^2          & 0      & 0      \\
		\sm{0}  & 0                                  & p^2q\!+\!q^3 & 0             & 2pq^2  & 0      \\
		\sm{1}  & \frac{1}{2}pq^2\!+\!\frac{1}{2}p^3 & 0            & 2p^2q\!+\!q^3 & 0      & 2pq^2  \\
		\sm{2}  & p^3\!+\!pq^2                       & 0            & 3p^2q\!+\!q^3 & 0      & 2pq^2
	} \IEEEeqnarraynumspace
\end{IEEEeqnarray}

\pagebreak
\section{Solving the recurrent equation}\label{sec:solving_the_recurrent_equation}
Consider the second equation in \eqref{eq: recurrent_equation}. It follows from \eqref{eq: viterbi_bit_error_probability_w} that we are only interested in the asymptotic values, and hence letting $t$ tend to infinity yields
\begin{IEEEeqnarray}{rCl}
	\label{eq: b_infinity}
	\bs{b}_{\infty} & = & \bs{b}_{\infty} \Pi
\end{IEEEeqnarray}
where $\bs{b}_{\infty}$ can be chosen as
\begin{IEEEeqnarray}{rCl}
\label{inf_case}
	\bs{b}_{\infty} & = & (\bs{p}_{\infty}\, \bs{p}_{\infty} \ldots \bs{p}_{\infty})
\end{IEEEeqnarray}
To obtain the last equality, we took into account that $\Pi$ is a block-diagonal matrix whose diagonal elements are given by the transition probability matrix $\Phi$ which satisfies \eqref{steady-state}. Based on these observations, \eqref{eq: recurrent_equation} can be simplified to
\begin{IEEEeqnarray}{rCl}
	\label{eq: simplified_recurrent_equation}
	\bs{w}_{t + 1} & = & \bs{w}_{t} A + \bs{b}_{\infty} B
\end{IEEEeqnarray}

By iterating the recurrent equation \eqref{eq: simplified_recurrent_equation} and using the initial value \eqref{eq: initial_value_w}, the vector of the information weights at time instant $t+1$ is given by
\begin{IEEEeqnarray}{rCl}
	\bs{w}_{t+1} & = & \bs{b}_{\infty} B A^t + \bs{b}_{\infty} B A^{t-1} + \cdots + \bs{b}_{\infty} B
	\label{eq: w_t_plus_1}
\end{IEEEeqnarray}
Taking its limit, it follows that
\begin{IEEEeqnarray}{rCl}
	\lim_{t \to \infty} \frac{\bs{w}_t}{tb} & = & \lim_{t \to \infty} \frac{\bs{w}_{t+1}}{tb} = \lim_{t \to \infty} \frac{1}{tb} \sum_{j=0}^t \bs{b}_{\infty} B A^{t-j} \nonumber \\
                                          & = & \bs{b}_{\infty} B A^{\infty}/b \label{eq: final_limit}
	\end{IEEEeqnarray}
where $A^{\infty}$ denotes the limit of the sequence $A^t$ when $t$ tends to infinity and we used the fact that, if a sequence converges to a finite limit, then it is Ces\`{a}ro-summable to the same limit.

From \eqref{eq:row_sum} it follows that
\begin{IEEEeqnarray}{rCl}
	\label{eq: left_eigenvector}
	\bs{e}_{\text{L}} & = & (\bs{p}_{\infty} \enspace \bs{p}_{\infty} \ldots \bs{p}_{\infty})
\end{IEEEeqnarray}
satisfies
\begin{IEEEeqnarray}{rCl}
	\bs{e}_{\text{L}} A & = & \bs{e}_{\text{L}}
\end{IEEEeqnarray}
and hence $\bs{e}_{\text{L}}$ is a left eigenvector with eigenvalue $\lambda = 1$. Due to the nonnegativity of $A$, $\lambda = 1$ is a maximal eigenvalue of $A$ (Corollary 8.1.30 \cite{Horn1990}). Let $\bs{e}_{\text{R}}$ denote the right eigenvector corresponding to the eigenvalue $\lambda = 1$ normalized such that $\bs{e}_{\text{L}} \bs{e}_{\text{R}} = 1$. If we remove the allzero rows and corresponding columns from the matrix $A$ we obtain an irreducible matrix which has a unique maximal eigenvalue $\lambda = 1$ (Lemma 8.4.3~\cite{Horn1990}). Hence, it follows (Lemma 8.2.7, statement (i)~\cite{Horn1990}) that
\begin{IEEEeqnarray}{rCl}
	\label{eq: multiplication_eigenvectors}
	A^{\infty} = \bs{e}_{\text{R}} \bs{e}_{\text{L}}
\end{IEEEeqnarray}
Combining \eqref{eq: final_limit}, \eqref{eq: left_eigenvector}, and \eqref{eq: multiplication_eigenvectors} yields
\begin{IEEEeqnarray}{rCl}
	\label{eq: combination}
	\lim_{t \to \infty} \frac{\bs{w}_t}{tb} & = & \bs{b}_{\infty} B \bs{e}_{\text{R}} ( \bs{p}_{\infty} \enspace \bs{p}_{\infty} \ldots \bs{p}_{\infty}) / b
\end{IEEEeqnarray}
Following \eqref{eq: viterbi_bit_error_probability_w}, by summing up the first $M$ components of the vector $( \bs{p}_{\infty} \enspace \bs{p}_{\infty} \ldots \bs{p}_{\infty})$ on the right side of \eqref{eq: combination}, we obtain the closed form expression for the exact bit error probability as
\begin{IEEEeqnarray}{rCl}
	\label{eq: final}
	P_{\text{b}} & = & \bs{b}_{\infty} B \bs{e}_{\text{R}} / b
\end{IEEEeqnarray}

To summarize, the exact bit error probability $P_\text{b}$ for Viterbi decoding of a rate $R=b/c$ minimal convolutional encoder, when communicating over the BSC, is calculated as follows:
\begin{itemize}
	\item Construct the set of metric states and find the stationary probability distribution $\bs{p}_{\infty}$.
	\item Determine the matrices $A$ and $B$ as in \secref{recurrent_equation} and compute the right eigenvector $\bs{e}_{\text{R}}$ normalized according to $(\bs{p}_{\infty} \enspace \bs{p}_{\infty} \ldots \bs{p}_{\infty}) \bs{e}_{\text{R}} = 1$.
	\item Calculate the exact bit error probability $P_{\text{b}}$ using \eqref{eq: final}.
\end{itemize}
For the encoder shown in \figref{ocf_realization} we obtain
\begin{IEEEeqnarray}{rCl}
  P_{\text{b}} & = & \left(4p - 2p^2 + 67p^3 - 320p^4 + 818p^5 - 936p^6 - 884p^7 \right. \nonumber \\
               &   & \,\,+ 5592p^8 - 11232p^9 + 13680p^{10} - 11008p^{11} \nonumber \\
               &   & \left.\,\,+ 5760p^{12} - 1792p^{13} + 256p^{14} \right) / \left(2 - 5p + 41p^2 \right. \nonumber\\
               &   & \,\, - 128p^3 + 360p^4 - 892p^5 + 1600p^6 - 1904p^7 \nonumber \\
               &   & \left.\,\,+ 1440p^8 - 640p^9 + 128p^{10} \right) \nonumber \\
               & = & 2p + 4p^2 + \frac{5}{2}p^3 - \frac{431}{4}p^4 - \frac{125}{8}p^5 + \frac{32541}{16}p^6 \nonumber \\
               &   & \, - \frac{70373}{32}p^7 - \frac{1675587}{64}p^8 + \frac{7590667}{128}p^9 \nonumber \\
               &   & + \frac{67672493}{256}p^{10} - \cdots
\end{IEEEeqnarray}

If we instead realize the \textit{minimal} generator matrix \eqref{eq:ocf_encoding_matrix_R23} in controller canonical form, we obtain a \textit{nonminimal} ($4$-state) encoder with $M=12$ normalized cumulative metric state; \textit{cf.},~the Remark after \eqref{eq: viterbi_bit_error_probability}. Its exact bit error probability is slightly worse than that of its minimal realization in observer canonical form.

\section{Some Examples}\label{sec:examples}

\begin{figure}[t]
	\centering
	\includegraphics{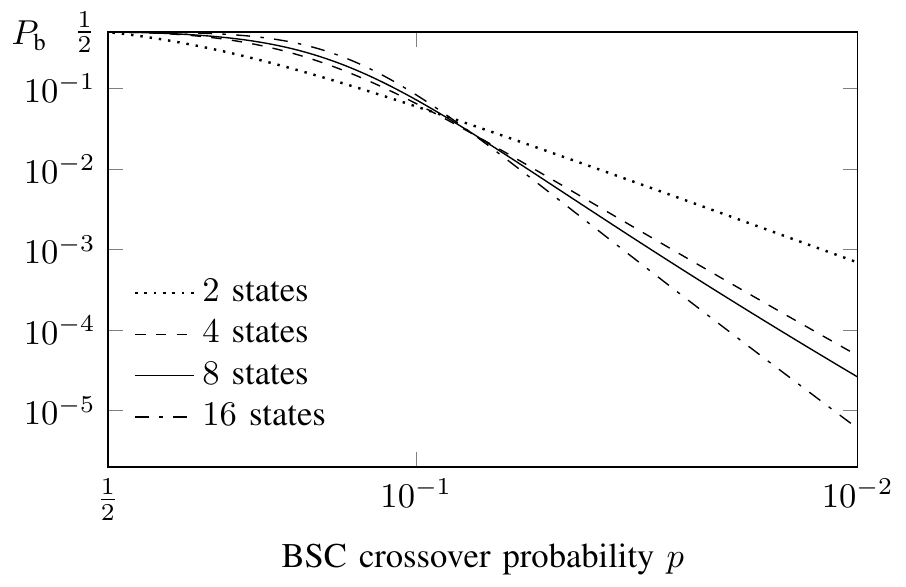}
	\vspace{-1em}
	\caption{\label{fig:graph}Exact bit error probability for the rate $R=1/2$ minimal encoders of memory $m=1$ ($2$-state) $G(D)=(1 \quad 1+D)$, memory $m=2$ ($4$-state) $G(D)=(1+D^2 \quad 1+D+D^2)$, memory $m=3$ ($8$-state) $G(D)=(1+D^2+D^3 \quad 1+D+D^2+D^3)$, and memory $m=4$ ($16$-state) $G(D)=(1+D+D^4 \quad 1+D+D^2+D^3+D^4)$.}
	\vspace{-1em}
\end{figure}

\begin{figure*}[t]
	\centering
  	\includegraphics{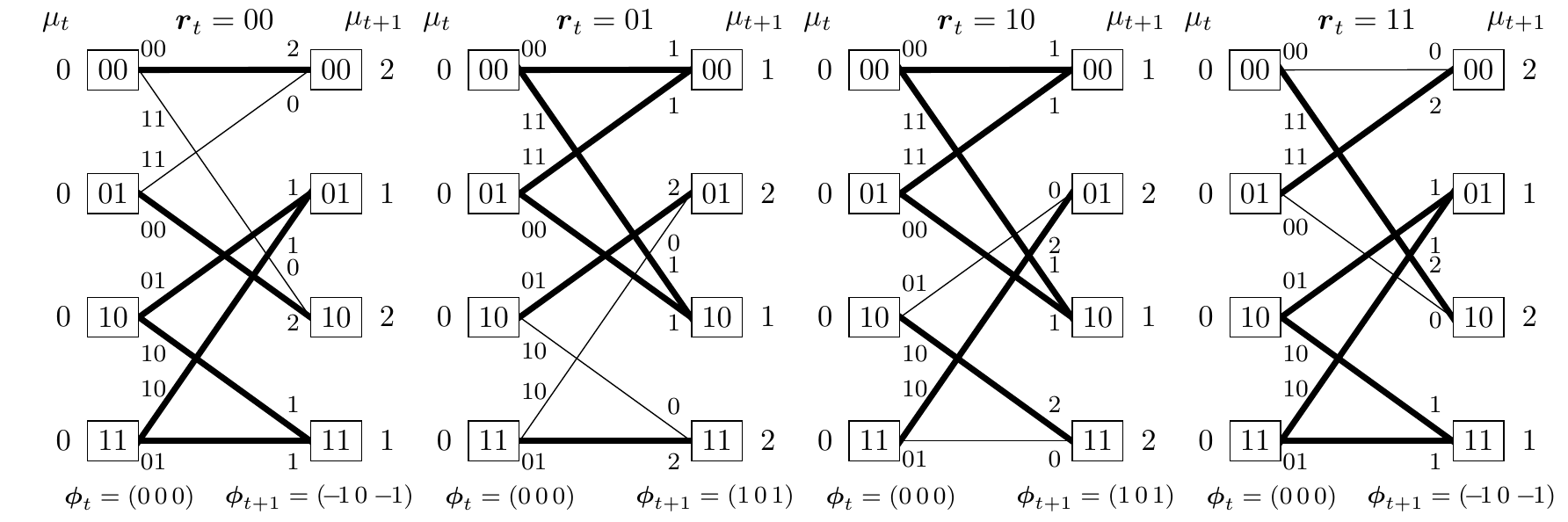}
	\caption{\label{fig:four_trellis_sections_m2}Four different trellis sections of the in total $124$ for the $G(D) = (1+D^2 \quad 1+D+D^2)$ generator matrix.}
	\vspace{-1em}
\end{figure*}

First we consider some rate $R=1/2$, memory $m=1,2,3$, and $4$ convolutional encoders; that is, encoders with $2, 4, 8$, and $16$ states, realized in controller canonical form. In \figref{graph} we plot the exact bit error probability for those four convolutional encoders.
\begin{example}
  If we draw all $20$ trellis sections for the rate $R=1/2$, memory $m=1$ ($2$-state) convolutional encoder with generator matrix $G(D) = (1 \quad 1+D)$ realized in controller canonical form, we obtain the normalized cumulative metric states $\left\{ -2, -1, 0, 1, 2 \right\}$. Its metric state Markov chain yields the stationary probability distribution
  {\setlength\arraycolsep{0.1em}
  \begin{equation}
   \bs{p}^\text{T}_{\infty} = \frac{1}{1+3p^2-2p^3}
   \left(\begin{array}{ccccccccc}
           1 & - & 4p & + & 8p^2 & - & 7p^3 & + & 2p^4 \\
             &   & 2p & - & 5p^2 & + & 5p^3 & - & 2p^4 \\
             &   & 2p & - & 3p^2 & + & 3p^3 \\
             &   &    &   & 2p^2 & - & 3p^3 & + & 2p^4 \\
             &   &    &   &  p^2 & + &  p^3 & - & 2p^4
    \end{array}\right) \label{eq:stationary_memory1}
  \end{equation}}%
  Based on these $20$ trellis sections, the $10 \times 10$ matrices $A$ and $B$ are constructed as
  \begin{equation}
    A = \left(\begin{array}{cc} A_{00} & A_{01} \\ A_{10} & A_{11} \end{array}\right)
  \end{equation}
  and
  \begin{equation}
    B = \left(\begin{array}{cc} \bs{0}_{5,5} & A_{01} \\ \bs{0}_{5,5} & A_{11} \end{array}\right) \label{eq:ex_B_matrix}
  \end{equation}
  where $\bs{0}_{5,5}$ denotes the $5 \times 5$ all-zero matrix. The normalized right eigenvector of $A$ is
  \begin{IEEEeqnarray}{rCl}
		\label{eq:right_eigenvector_memory_1}
		\bs{e}_{\text{R}} & = & \scriptscriptstyle
		\left( \begin{array}{c}
			0 \\ 0 \\ 0 \\ 0 \\ 0 \\ 0 \\ \displaystyle\frac{pq}{2} \\[0.7em] \displaystyle\frac{4pq}{2-p+4p^2-4p^3} \\[1em] \displaystyle\frac{(2 + 7p - 12p^2 + 13p^3 - 12p^4 + 4p^5)}{2(2-p+4p^2-4p^3)} \\[0.7em] 1
		\end{array}\right)
	\end{IEEEeqnarray}
	Finally, inserting \eqref{eq:stationary_memory1}, \eqref{eq:ex_B_matrix}, and \eqref{eq:right_eigenvector_memory_1} into \eqref{eq: final} yields the following expression for the exact bit error probability
	\begin{IEEEeqnarray}{rCl}
		P_{\text{b}} & = & \frac{14p^2 - 23p^3 + 16p^4 + 2p^5 - 16p^6 + 8p^7}{(1+3p^2-2p^3)(2 - p + 4p^2 - 4p^3)} \nonumber \\
		             & = & 7p^2 - 8p^3 - 31p^4 + 64p^5 + 86p^6 - \frac{635}{2}p^7 \nonumber \\
		             &   & \,- \frac{511}{4}p^8 + \frac{10165}{8}p^9 - \frac{4963}{16}p^{10} - \cdots
	\end{IEEEeqnarray}
	which coincides with the exact bit error probability formula given in \cite{Best1995}.
\end{example}

\begin{example}
  For the rate $R=1/2$, memory $m=2$ ($4$-state) convolutional encoder with generator matrix $G(D) = (1+D^2 \quad 1+D+D^2)$ realized in controller canonical form, we obtain, for example, the four trellis sections for $\phi_t = \left( 000 \right)$ shown in \figref{four_trellis_sections_m2}. The corresponding metric states at times $t+1$ are $\phi_{t+1} = \left( -1\,0\,-\!\!1\right)$ and $\phi_{t+1} = \left( 1\,0\,1 \right)$.

  Completing the set of trellis sections yields in total $M=31$ different normalized cumulative metric states, and hence the $124 \times 124$ matrices $A$ and $B$ have the following block structure
  \pagebreak
  \vphantom{.}\\\vspace{-0.5em}
  \begin{equation}
		A =
		\left(\begin{array}{cccc}
			A_{00} & \bs{0}_{31,31} & A_{02} & \bs{0}_{31,31} \\
			A_{10} & \bs{0}_{31,31} & A_{12} & \bs{0}_{31,31} \\
			\bs{0}_{31,31} & A_{21} & \bs{0}_{31,31} & A_{23} \\
			\bs{0}_{31,31} & A_{31} & \bs{0}_{31,31} & A_{33}
		\end{array}\right)
	\end{equation}
	\vspace{-0.5em}
	and
	\begin{equation}
		B =
		\left(\begin{array}{cccc}
			\bs{0}_{31,31} & \bs{0}_{31,31} & A_{02} & \bs{0}_{31,31} \\
			\bs{0}_{31,31} & \bs{0}_{31,31} & A_{12} & \bs{0}_{31,31} \\
			\bs{0}_{31,31} & \bs{0}_{31,31} & \bs{0}_{31,31} & A_{23}\\
			\bs{0}_{31,31} & \bs{0}_{31,31} & \bs{0}_{31,31} & A_{33}
		\end{array}\right)
  \end{equation}

  Following the method for calculating the exact bit error probability described in \secref{solving_the_recurrent_equation} we obtain
	\begin{IEEEeqnarray}{rCl}
		P_{\text{b}} & = & 44p^3 + \frac{3519}{8}p^4 - \frac{14351}{32}p^5 - \frac{1267079}{64}p^6 \nonumber\\
		             &   & \,- \frac{31646405}{512}p^7 + \frac{978265739}{2048}p^8 \nonumber\\
		             &   & \, + \frac{3931764263}{1024}p^9 - \frac{48978857681}{32768}p^{10} - \cdots \label{eq:ebep_m2}
	\end{IEEEeqnarray}
	which coincides with the previously obtained result by Lentmaier et al.~\cite{Lentmaier2004}.
\end{example}

\begin{example}
  For the rate $R=1/2$, memory $m=3$ ($8$-state) convolutional encoder with generator matrix $G(D) = (1+D^2+D^3 \quad 1+D+D^2+D^3)$ realized in controller canonical form we have $M=433$ normalized cumulative metric states and the $A$ and $B$ matrices are of size $433\cdot 2^3 \times 433 \cdot 2^3$.

  Since the complexity of the symbolic derivations increases greatly, we can only obtain a numerical solution of \eqref{eq: final}, as shown in \figref{graph}.
\end{example}

\begin{example}
  For the rate $R=1/2$, memory $m=4$ ($16$-state) convolutional encoder with generator matrix $G(D) = (1+D^2+D^3+D^4 \quad 1+D+D^4)$ realized in controller canonical form, we have as many as $M=188687$ normalized cumulative metric states. Thus, the matrices $A$ and $B$ are of size $188687 \cdot 2^4 \times 188687 \cdot 2^4$. The corresponding numerical solution of \eqref{eq: final} is plotted in \figref{graph}.
\end{example}

The obvious next step is to try a rate $R=1/2$, memory $m=5$ ($32$-state) convolutional encoder. We tried the generator matrix $G(D) = (1+D+D^2+D^3+D^4+D^5 \quad 1+D^3+D^5)$ realized in controller canonical form but were only able to show that the number of cumulative normalized metric states $M$ exceeds $4130000$.

\begin{figure}[!b]
	\centering
	\vspace{-1em}
	\includegraphics{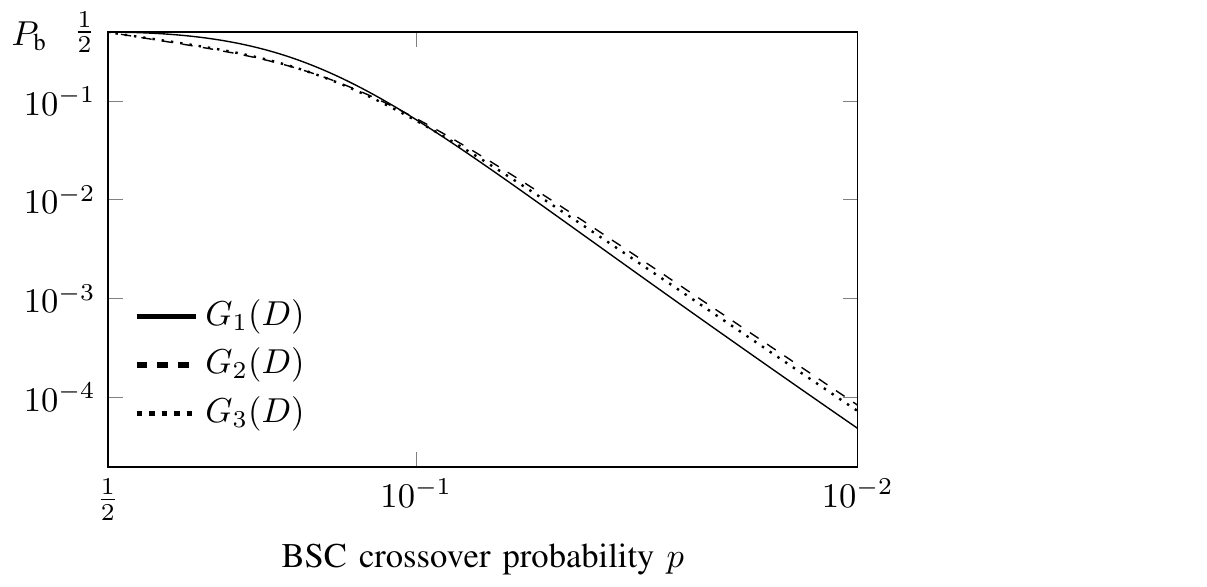}
	\vspace{-1.5em}
	\caption{\label{fig:equivalent_m2}Exact bit error probability for the rate $R=1/2$ memory $m=2$ minimal encoders with $G_1(D) = ( 1+D^2 \quad 1+D+D^2)$, $G_2(D) = (1 \quad (1+D^2)/(1+D+D^2))$, and $G_3(D) = (1 \quad (1+D+D^2)/(1+D^2))$.}
\end{figure}

\begin{example}
  Consider the generator matrix $G_1(D) = ( 1+D^2 \quad 1+D+D^2 )$ and its equivalent systematic generator matrices $G_2(D) = (1 \quad (1+D^2)/(1+D+D^2))$ and $G_3(D) = (1 \quad (1+D+D^2)/(1+D^2))$. When realized in controller canonical form, all three realizations have $M=31$ normalized cumulative metric states. The exact bit error probability for $G_1(D)$ is given by \eqref{eq:ebep_m2}. For $G_2(D)$ and $G_3(D)$ we obtain
  \begin{IEEEeqnarray}{rCl}
		P_{\text{b}} & = & \frac{163}{2}p^3 + \frac{365}{2}p^4 - \frac{24045}{8}p^5 - \frac{1557571}{128}p^6 \nonumber\\
		             &   & \,+ \frac{23008183}{512}p^7 + \frac{1191386637}{2048}p^8 \nonumber \\
		             &   & \,+ \frac{4249634709}{8192}p^9 + \frac{132555764497}{8192}p^{10} - \cdots \label{eq:ebep_m2_5_7}
	\end{IEEEeqnarray}
	and\vspace{-0.5em}
	\begin{IEEEeqnarray}{rCl}
		P_{\text{b}} & = & \frac{141}{2}p^3 + \frac{1739}{8}p^4 - \frac{71899}{32}p^5 - \frac{1717003}{128}p^6 \nonumber \\
		             &   & \,+ \frac{2635041}{128}p^7 + \frac{540374847}{1024}p^8 \nonumber \\
		             &   & \,+ \frac{9896230051}{8192}p^9 - \frac{402578056909}{32768}p^{10}	- \cdots \label{eq:ebep_m2_7_5}
	\end{IEEEeqnarray}
	respectively. The corresponding numerical results are illustrated in \figref{equivalent_m2}.
\end{example}

\begin{figure}[t]
	\centering
	\includegraphics{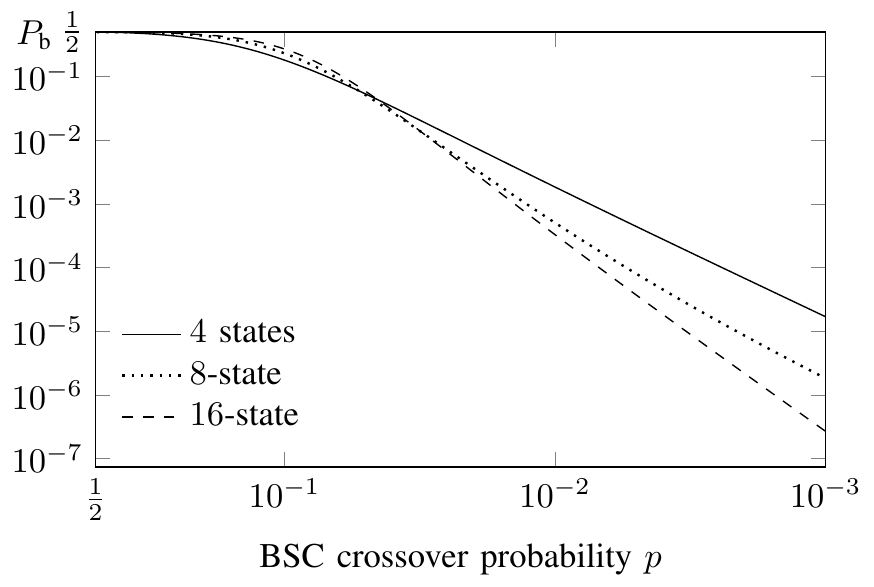}
	\vspace{-1em}
	\caption{\label{fig:ebep_r23}Exact bit error probability for the rate $R=2/3$, overall constraint length $\nu=2,3$, and $4$ ($4$-state, $8$-state, and $16$-state, respectively) minimal encoders whose generator matrices are given in \tabref{encodingmatrices}.}
\end{figure}

\begin{example}
  The exact bit error probabilities for the rate $R=2/3$ $4$-state, $8$-state, and $16$-state generator matrices, given in \tabref{encodingmatrices} and realized in controller canonical form, are plotted in \figref{ebep_r23}.

  \begin{table}[!t]
    \centering
    \caption{\label{tab:encodingmatrices}Rate $R=2/3$ generator matrices}
    \renewcommand{\arraystretch}{1.2}
    \setlength\tabcolsep{0.2em}
    \begin{tabular}{c|c|c|c}
    $G(D)$ & \#states & $d_\text{free}$ & $M$ \\[0.5em] \shhline[1pt]
    \multirow{3}{*}{\renewcommand{\arraystretch}{1}$\left(\begin{array}{ccc} D & 1+D & 1+D \\ 1 & D & 1+D \end{array}\right)$            } &  \multirow{3}{*}{$4$} &\multirow{3}{*}{$3$} &     \multirow{3}{*}{$19$} \\ &&\\ &&\\ \shhline
    \multirow{3}{*}{\renewcommand{\arraystretch}{1}$\left(\begin{array}{ccc} 1+D & D & 1 \\ D^2 & 1 & 1+D+D^2 \end{array}\right)$        } &  \multirow{3}{*}{$8$} &\multirow{3}{*}{$4$} &    \multirow{3}{*}{$347$} \\ &&\\ &&\\ \shhline
    \multirow{3}{*}{\renewcommand{\arraystretch}{1}$\left(\begin{array}{ccc} D+D^2 & 1 & 1+D^2 \\ 1 & D+D^2 & 1+D+D^2 \end{array}\right)$} & \multirow{3}{*}{$16$} &\multirow{3}{*}{$5$} &  \multirow{3}{*}{$15867$} \\ &&\\ &&\\ \shhline[1pt]
    \end{tabular}\vspace{0.5em}
  \end{table}

  As an example, the $4$-state encoder has the exact bit error probability
  \begin{IEEEeqnarray}{rCl}
		P_{\text{b}} & = & \frac{67}{2}p^2+ \frac{17761}{48}p^3 - \frac{2147069}{648}p^4 - \frac{1055513863}{46656}p^5 \nonumber\\
		             &   & \,+ \frac{123829521991}{559872}p^6 + \frac{67343848419229}{60466176}p^7 \nonumber\\
		             &   & \,- \frac{27081094434882419}{2176782336}p^8 - \frac{477727138796620247}{8707129344}p^9 \nonumber \\
		             &   & \,+ \frac{1944829319763332473469}{2821109907456}p^{10} + \cdots
	\end{IEEEeqnarray}
\end{example}

If we replace the BSC with the quantized additive white Gaussian noise (AWGN) channel, the calculation of the exact bit error probability follows the same method as described in \secref{solving_the_recurrent_equation}, but the computational complexity increases dramatically as illustrated by the following example.

\begin{figure}[t]
	\centering
	\includegraphics{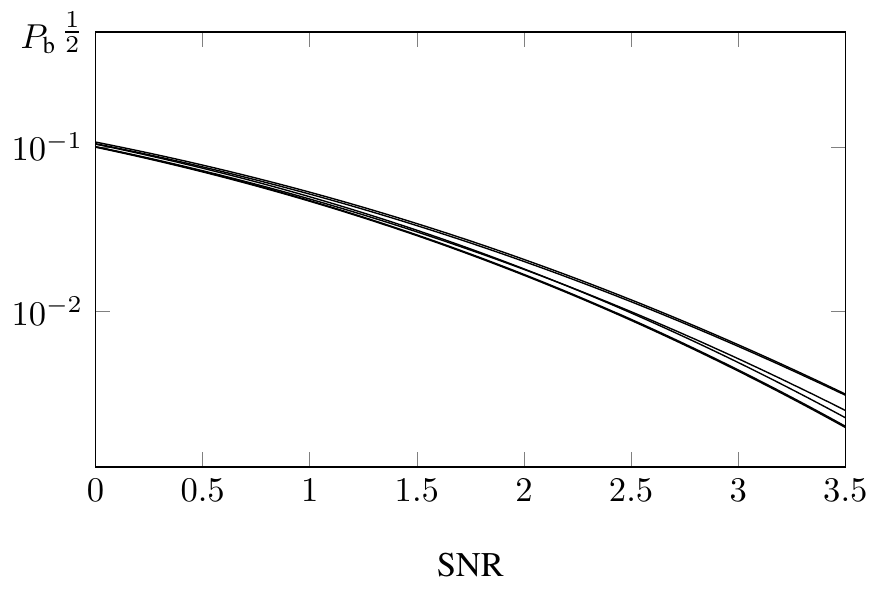}
	\vspace{-1em}
	\caption{\label{fig:ebep_quant}Exact bit error probability for the rate $R=1/2$, memory $m=2$ ($4$-state) encoder with $G(D)=(1+D^2 \quad 1+D+D^2)$ used to communicate over an AWGN channel with different quantization levels.}
	\vspace{-0.8em}
\end{figure}

\begin{figure}[ht]
	\centering
	\includegraphics{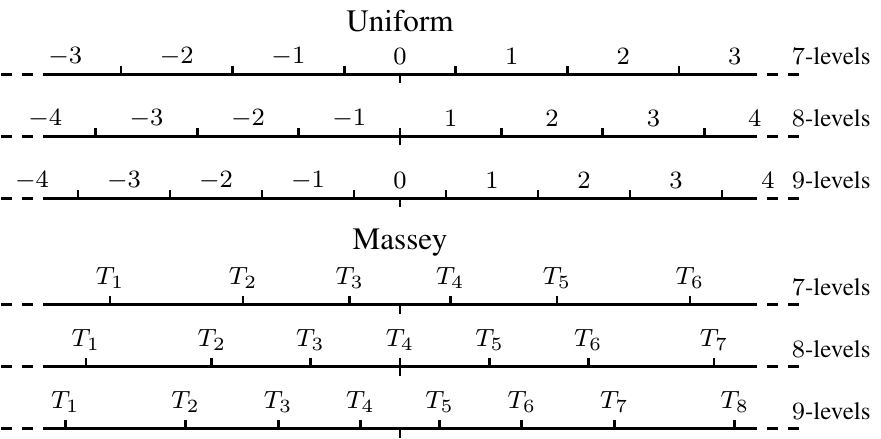}
	\vspace*{1mm}
	\caption{\label{fig:quant_levels}Examples of uniform and Massey quantizations for an AWGN channel with SNR = $0 \text{dB}$.}
	\vspace{-1em}
\end{figure}

\begin{example}
  Consider the generator matrix $G(D) = (1+D^2 \quad 1+D+D^2)$ used to communicate over a quantized AWGN channel. We use different quantization methods, namely, uniform quantization \cite{Heller1971, Onyszchuk1993} and Massey quantization \cite{Massey1974, Johannesson1999}; see \figref{quant_levels}.

  The uniform intervals were determined by optimizing the cut-off rate $R_0$. The Massey quantization thresholds $T_i$ between intervals were also determined by optimizing $R_0$, but allowing for nonuniform intervals. The realization in controller canonical form yields that, for all signal to noise ratios (SNRs), $E_\text{b} / N_0$, and uniform quantization with $7$, $8$, and $9$ levels, the number of the normalized cumulative metric states is $M=1013$, $M=2143$, and $M=2281$, respectively. However, for the Massey quantization the number of normalized cumulative metric states varies with both the number of levels and the SNR. Moreover, these numbers are much higher. For example, considering the interval between $0\,\text{dB}$ and $3.5\,\text{dB}$ with $8$ quantization levels, we have $M=16639$ for $E_\text{b} / N_0 \leq 2.43\,\text{dB}$, while for $E_\text{b} / N_0 > 2.43\,\text{dB}$ we obtain $M=17019$. The exact bit error probability for this $4$-state encoder is plotted for all different quantizations in \figref{ebep_quant}, ordered from worst (top) to best (bottom) as
  
  {\vspace{0.5em}\setlength\tabcolsep{0em}
  \begin{tabular}{rll}
    \quad (i)   &\, Uniform~~& $8$ levels \\
    \quad (ii)  &\, Uniform~~& $7$ levels \\
    \quad (iii) &\, Massey ~~& $7$ levels \\
    \quad (iv)  &\, Uniform~~& $9$ levels \\
    \quad (v)   &\, Massey ~~& $8$ levels \\
    \quad (vi)  &\, Massey ~~& $9$ levels \\
  \end{tabular}\vspace{0.5em}}%
  
  All differences are very small, and hence it is hard to distinguish all the curves. It is interesting to notice that using $7$ instead of $8$ uniform quantization levels yields a better bit error probability. However, this is not surprising since the presence of a quantization bin around zero typically improves the quantization performance. Moreover, the number of cumulative normalized metric states for $7$ quantization levels is only about one half of that for $8$ quantization levels. Notice that such a subtle comparison of channel output quantizers has only become possible due to the closed form expression for the exact bit error probability.
\end{example}

\section{Conclusion}
We have derived a closed form expression for the exact bit error probability for Viterbi decoding of convolutional codes using a recurrent matrix equation. In particular, the described method is feasible to evaluate the performance of encoders with as many as $16$ states when communicating over the BSC. By applying our new approach to a $4$-state encoder used to communicate over the quantized AWGN channel, the expression for the exact error probability for Viterbi decoding is also derived. In particular, it is shown that the proposed technique can be used to select the optimal encoder implementation as well as the optimal channel output quantizer based on comparing their corresponding exact bit decoding error probability.


\vfill
\pagebreak

\begin{IEEEbiographynophoto}{Irina E. Bocharova}
was born in Leningrad, U.S.S.R., on July 31, 1955. She received the Diploma in Electrical Engineering in 1978 from the Leningrad Electro-Technical Institute and the Ph.D. degree in technical sciences in 1986 from the Leningrad Institute of Aerospace Instrumentation.

During 1986-2007, she has been a Senior Researcher, an Assistant Professor, and then Associate Professor at the Leningrad Institute of Aerospace Instrumentation (now State University of Aerospace Instrumentation, St.-Petersburg, Russia). Since 2007 she has been an Associate Professor at the State University of Information Technologies, Mechanics and Optics. Her research interests include convolutional codes, communication systems, source coding and its applications to speech, audio and image coding. She has published more than 50 papers in journals and proceedings of international conferences, and seven U.S. patents in speech, audio and video coding. She has authored the textbook \textit{Compression for Multimedia} (Cambridge University Press, 2010).

Professor Bocharova was awarded the Lise Meitner Visiting Chair in engineering at Lund University, Lund, Sweden (January--June 2005 and 2011).  
\end{IEEEbiographynophoto}

\begin{IEEEbiographynophoto}{Florian Hug}
(S'08) was born in Memmingen, Germany, on May 21, 1983. He received the Dipl.-Ing. degree from the Faculty of Electrical Engineering, University of Ulm, Ulm, Germany, in 2008.
Since then, he has been with the Department of Electrical and Information Technology, Lund University, Lund, Sweden, where he is working towards the Ph.D. degree in information theory. His research interests covers the field of information and coding theory. Currently, he is focusing on codes over graphs.
\end{IEEEbiographynophoto}
\begin{IEEEbiographynophoto}{Rolf Johannesson}
(M'72, F'98, LF'12) was born in H\"assleholm, Sweden, on July 3, 1946. He received the M.S. and Ph.D. degrees in 1970 and 1975, respectively, both from Lund University, Lund, Sweden, and the degree of Professor, \textit{honoris causa}, from the Institute for Problems of Information Transmission, Russian Academy of Sciences, Moscow, in 2000. 

Since 1976, he has been a faculty member with Lund University where he has the Chair of Information Theory. From 1976 to 2003, he was department Head and during 1988-1993 Dean of the School of Electrical Engineering and Computer Sciences. During 1990-1995, he served as a member of the Swedish Research Council for Engineering Sciences. His scientific interests include information theory, error-correcting codes, and cryptography. In addition to papers in the area of convolutional codes and cryptography, he has authored two textbooks on switching theory and digital design (both in Swedish) and one on information theory (in both Swedish and German) and coauthored \textit{Fundamentals of Convolutional Coding} (New York: IEEE Press, 1999), and \textit{Understanding Information Transmission} (Hoboken, NJ: IEEE Press/Wiley-Interscience, 2005).

Professor Johannesson has been an Associate Editor for the \textit{International Journal of Electronics and Communications}. During 1983-1995 he co-chaired seven Russian-Swedish Workshops, which were the chief interactions between Russian and Western scientists in information theory and coding during the final years of the Cold War. He became an elected member of the Royal Swedish Academy of Engineering Sciences in 2006.
\end{IEEEbiographynophoto}
\begin{IEEEbiographynophoto}{Boris D. Kudryashov}
was born in Leningrad, U.S.S.R., on July 9, 1952. He received the Diploma in Electrical Engineering in 1974 and the Ph.D in technical sciences degree in 1978 both from the Leningrad Institute of Aerospace Instrumentation, and the Doctor of Science degree in 2005 from Institute of Problems of Information Transmission, Moscow. 

In 1978, he became an Assistant Professor and then Associate Professor and Professor in the Leningrad Institute of Aerospace Instrumentation (now the State University on Aerospace Instrumentation, St.-Petersburg, Russia). Since November 2007, he has been a Professor with the State University on Information Technologies, Mechanics and Optics, St.-Petersburg, Russia. His research interests include coding theory, information theory and applications to speech, audio and image coding. He has authored a textbook on information theory (in Russian) and has more than 70 papers published in journals and proceedings of international conferences, 15 U.S. patents and published patent applications in image, speech and audio coding. 

Professor Kudryashov served as a member of Organizing Committee of ACCT International Workshops.
\end{IEEEbiographynophoto}
\vfill
\end{document}